\documentclass[final,5p,times,twocolumn]{elsarticle}

\usepackage{graphicx}

\usepackage{amssymb}

\journal{Journal of Physics and Chemistry of Solids}

\begin{document}

\begin{frontmatter}



\title{Large upper critical field in non-centrosymmetric superconductor Y$_2$C$_3$}


\author[1]{H. Q. Yuan\corref{cor1}}
\cortext[cor1]{Corresponding author.}
\ead{hqyuan@zju.edu.cn}
\author[1]{J. Chen}
\author[2]{J. Singleton}
\author[3]{S. Akutagawa}
\author[3]{J. Akimitsu}

\address[1]{Department of Physics, Zhejiang University, Hangzhou 310027, P. R. China }
\address[2]{NHMFL, Los Alamos National Laboratory, MS E536, Los Alamos, NM 87545, USA.}
\address[3]{Department of Physics and Mathematics, Aoyama Gakuin University,Sagamihara, Kanagawa
229-8558, Japan}

\begin{abstract}

We determine the upper critical field $\mu_0H_{c2}(T_c)$ of
non-centrosymmetric superconductor Y$_2$C$_3$ using two distinct
methods: the bulk magnetization $M(T)$ and the tunnel-diode
oscillator (TDO) based impedance measurements. It is found that the
upper critical field reaches a value of 30T at zero temperature
which is above the weak-coupling Pauli paramagnetic limit. We argue
that the observation of such a large $\mu_0H_{c2}(0)$ in Y$_2$C$_3$
could be attributed to the admixture of spin-singlet and
spin-triplet pairing states as a result of broken inversion
symmetry.

\end{abstract}


\end{frontmatter}



\section{Introduction}
\label{}

Among the previously investigated superconductors, most of them
posses a center of inversion in their crystal structure. In these
cases, the Pauli principle and the parity conservation then dictate
that the superconducting pairing states with even parity are
necessarily spin singlet, while those with odd parity must be spin
triplet \cite{Anderson59, Anderson84}. However, such a tie between
the spatial symmetry and the Cooper-pair spin may be broken in the
non-centrosymmetric superconductors in which the inversion symmetry
is absent \cite{Gor'kov, Yip, Frigeri04, Samokhin}.

The recent discovery of superconductivity in CePt$_3$Si
\cite{Bauer04} has largely stimulated the investigation of
superconductivity lacking inversion symmetry. In the past few years,
a growing number of non-centrosymmetric superconductors have been
successfully synthesized. Evidence for the admixture of spin-singlet
and spin-triplet components has been found in  Li$_2$(Pd,Pt)$_3$B
\cite{Yuan, Yuan 08, Nishiyama, Takeya} and the heavy fermion
systems of CePt$_3$Si \cite{Bauer07,Takeuchi,Bonalde}, CeRhSi$_3$
\cite{Kimura76, Kimura98} and CeIrSi$_3$ \cite{Sugitani,Okuda,
Mukuda}. In Li$_2$(Pd,Pt)$_3$B, it was shown that the spin triplet
state might become dominant with enhancing the antisymmetric
spin-orbit coupling (ASOC) strength \cite{Yuan, Yuan 08}. On the
other hand, the superconducting pairing state might become
complicated by strong electron correlations in the heavy fermion
compounds. Therefore, materials having weak electronic correlations
are desired for studying the effect of ASOC on superconductivity.

Y$_2$C$_3$ is a non-centrosymmetric superconductor with a
superconducting transition temperature $T_c$ of up to 18K, the
highest among all the investigated non-centrosymmetric
superconductors \cite{Amano}. It crystalizes in a body-centered
cubic structure (Pu$_2$C$_3$-type) with space group
I$\overline{4}$3d. Until now, only a very few physical properties
have been studied for Y$_2$C$_3$. Measurements of specific heat
indicates an isotropic superconducting gap of s-wave type
\cite{Akutagawa76}, but the NMR \cite{Harada} and $\mu$SR
\cite{Kuroiwa} experiments support a scenario of muti-gap
superconductivity. The upper critical field $\mu_{0}H_{c2}(0)$
derived from the initial slope at $T_c$ is estimated to be over 30T
\cite{Nakane04}, which is close to the Pauli paramagnetic limit.
Such a large upper critical field might indicate the significant
contribution of spin triplet component to the pairing state. To
determine the true value of the upper critical field and its
temperature dependence, one needs extend measurements to much higher
magnetic fields.

In this article, we report the upper critical field
$\mu_0H_{c2}(T_c)$ of Y$_2$C$_3$ measured by magnetization $M(T)$ up
to 14T and by the TDO resonant circuit using the facilities of
pulsed magnetic field at Los Alamos. An upper critical field of
about 30T is obtained with $T_c$ of 15.5K. We argue that the
contributions of spin triplet state resulting from the parity
violated ASOC might lead to the enhancement of the upper critical
field.

\section{Experimental methods}

Polycrystalline samples were prepared by mixing appropriate amounts
of Y (99.9\%) and C (99.95\%)  powders in a dry box
\cite{Akutagawa76}. The binary alloys were first prepared by arc
melting, after which they were subjected to high-pressure and
high-temperature treatment to produce the sesquicarbide phase. All
subsequent sample treatments, including preparations for
high-pressure experiments, were performed in a dry helium
atmosphere. The melted samples were heated to 1400 - 1600 $^\circ$C
\ in 5 minutes and then maintained at this temperature for 30
minutes under a pressure of 4.0 - 5.0 GPa using a cubic-anvil
high-pressure apparatus. After that, the samples were quickly
quenched to room temperature within a few seconds. The achieved
ingots were examined by powder x-ray diffraction using a
conventional X-ray spectrometer with a graphite monochromator, which
confirmed Y$_2$C$_3$ as a single phase.

Since the samples of Y$_2$C$_3$ are air sensitive, one needs avoid
exposing the samples in air. In order to minimize the time for
setting up the experiments, we choose the contactless measurements
of magnetization and the TDO resonant frequency (see below) to
determine the upper critical field. The magnetization was measured
in a commercial Quantum Design magnetic properties measurement
system (MPMS) with a magnet field up to 14T. In order to measure
$\mu_0H_{c2}$ down to zero temperature, we performed experiments in
pulsed magnetic fields up to 35T at Los Alamos, using a TDO-based
technique in which two small counter-wound coils form the inductance
of a resonant circuit. The resonant frequency (about 31M Hz) can
depend on both the skin-depth (or, in the superconducting state, the
penetration depth) and the differential magnetic susceptibility of
the sample. A piece of thin sample (about $0.8\times0.6\times0.1$
mm$^3$) was inserted in the coils, which were immersed in $^3$He
liquid or $^3$He exchange gas, temperatures being measured with a
Cernox thermometer 5 mm away from the sample.

\section{Results and Discussion}

 \begin{figure}[htbp]
 \includegraphics[width=9cm]{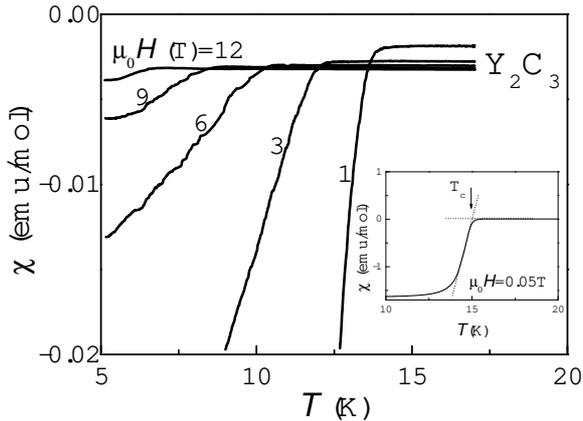}
 \caption{Temperature dependence of the magnetic susceptibility $\chi(T)$ at variant magnetic fields.
 $T_c$ is determined from the onset of the superconducting transition as shown in the inset.}\label{fig1}
 \end{figure}

Fig.1 presents the temperature dependence of the magnetic
susceptibility $\chi(T)$¡¡of Y$_2$C$_3$ at selected magnetic field.
To clearly show the superconducting transition at all the magnetic
fields, we only plot the superconducting transition partially for
the small fields. A rather sharp superconducting transition with
 $T_c\simeq15$K is observed at low field (see the inset), indicating good sample
quality. With increasing magnetic field, the superconducting
transition is shifted to lower temperature, but superconductivity is
not yet suppressed at a field as high as 14T. Therefore, facilities
with higher accessible magnetic field, e.g., the pulsed magnet, is
desired in order to track the critical field down to zero
temperature. From Fig. 1, one can find that the superconducting
transition becomes broader and the diamagnetic signal is weakened
upon applying a magnetic field which is likely attributed to the
flux pinning effect. In the context here, the superconducting
transition temperatures $T_c^{\chi}$ are determined from the onset
of the superconducting transition as shown in the inset of Fig. 1.

In order to further suppress superconductivity down to zero
temperature, we have measured the TDO resonant frequency of
Y$_2$C$_3$ up to 35T using a short pulse magnet. Fig. 2 shows the
relative TDO frequency, $\Delta f(\mu_0H)$, as a function of
magnetic field at selected temperatures for Y$_2$C$_3$. The sudden
increase of $\Delta f(\mu_0H)$ upon cooling down marks the onset of
the superconducting transition. At temperatures above $T_c$
($\simeq15$K), the TDO frequency increases smoothly with decreasing
temperature. The critical field of $\mu_0H_{c2}^{TDO}$ is determined
from the onset of the transition as described in the inset of Fig.
2. A magnetic field of about 30T is required to completely suppress
superconductivity in Y$_2$C$_3$ as seen in Fig.2. The successful
measurement of superconductivity and its upper critical field
$\mu_0H_{c2}(T_c)$ in Y$_2$C$_3$ indicates that the TDO-based
impedance measurement is a very powerful technique in studying
air-sensitive materials.

\begin{figure}[htbp]
 \includegraphics[width=9cm]{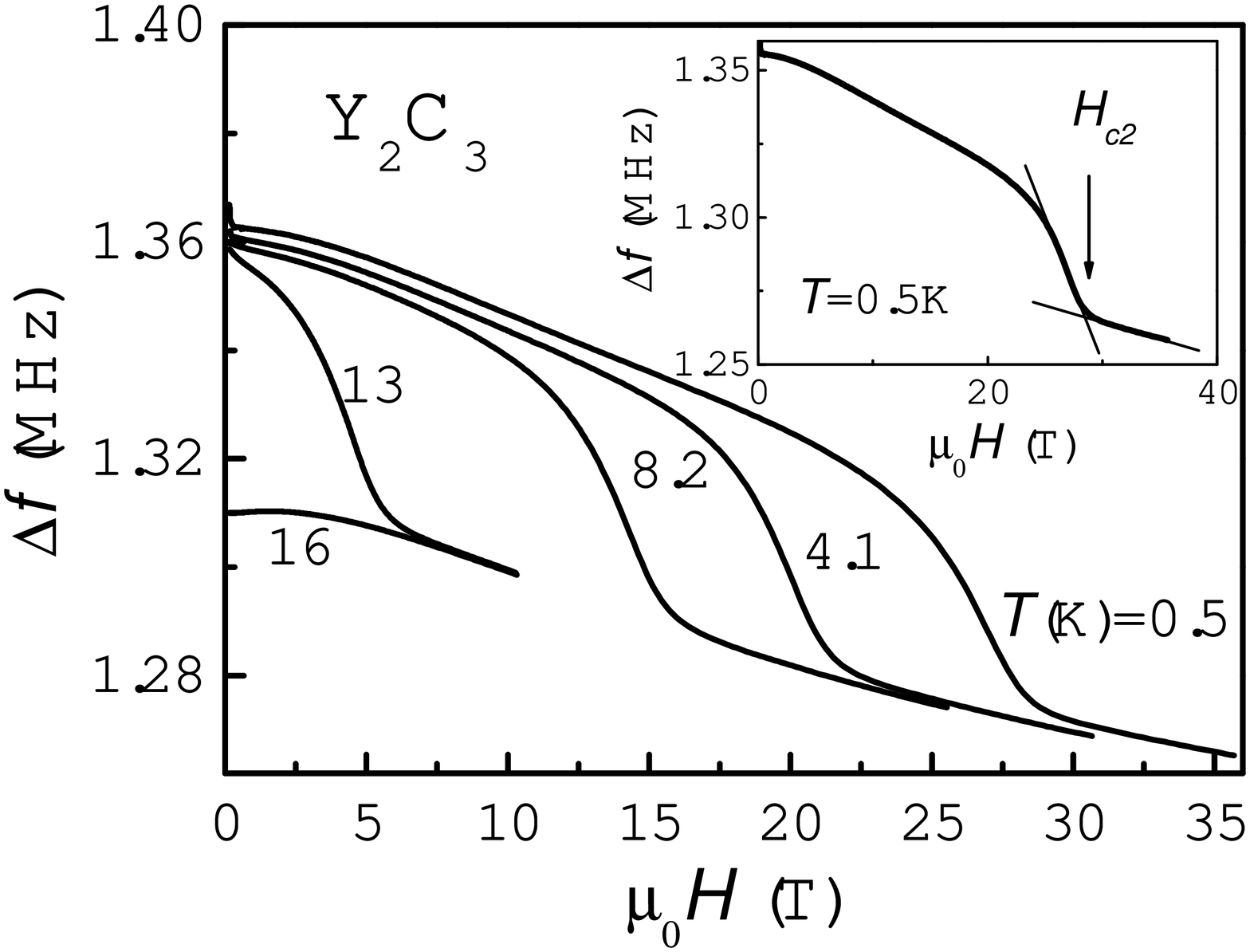}
 \caption{Magnetic field dependence of the resonant frequency shift $\Delta f(H)$
  at temperatures of $T=$0.5K, 4.1K, 8.2K, 13K and 16K. The inset shows $\Delta f(H)$ at $T=0.5$K, in which the
  critical field $\mu_0H_{c2}$ is determined from the intersection point.}\label{fig2}
 \end{figure}

The upper critical fields $\mu_0H_{c2}(T_c)$, obtained from the
magnetic susceptibility $\chi(T)$ and the TDO resonant frequency
$\Delta f(H)$ as described above, are plotted as a function of
temperature in Fig.3. The discrepancy between
$\mu_0H_{c2}^\chi(T_c)$ and $\mu_0H_{c2}^{TDO}(T_c)$ might originate
from the fact that the magnetic susceptibility measures the bulk
superconductivity while the TDO resonant technique is a kind of
surface measurement. Nevertheless, the two curves of
$\mu_0H_{c2}(T_c)$ follow qualitatively similar behavior:
$\mu_0H_{c2}$ shows a weak upturn close to $T_c$, and then increases
linearly with decreasing temperature. The extension of
$\mu_0H_{c2}(T_c)$ to zero gives $T_c^\chi$=15K and
$T_c^{TDO}$=15.5K, respectively. Two important features can be
observed in Fig. 3: (i) Y$_2$C$_3$ posses a large value of upper
critical field ($\mu_0H_{c2}^{\chi}(0)\simeq 30$ T) and (ii)
$\mu_0H_{c2}(T_c)$ shows very unusual temperature dependence.

 \begin{figure}[htbp]
 \includegraphics[width=9cm]{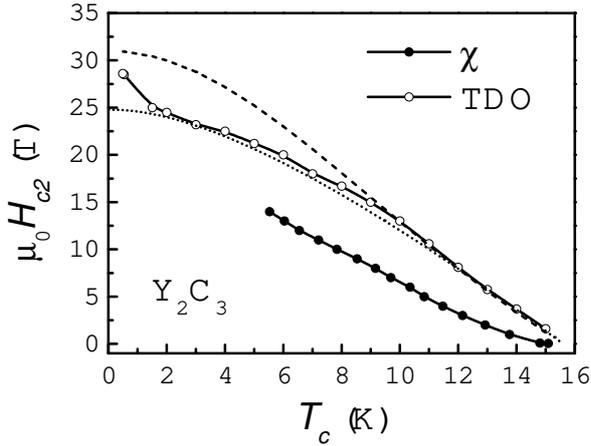}
 \caption{The upper critical field versus temperature for Y$_2$C$_3$. The filled and open symbols
 represent the data derived from the magnetization and the relative TDO resonant frequency, respectively. The dotted line and the dashed line are fits to the weak coupling WHH theory and $\mu_0H_{c2}(T_c)=\mu_0H_{c2}(0)\frac{1-(T/T_c)^2}{1+(T/T_c)^2}$, respectively.}\label{fig3}
 \end{figure}

In a conventional BCS superconductor, two electrons with opposite
spins and momenta form a Cooper pair. Application of a magnetic
field can break the Cooper pairs and, therefore, suppress
superconductivity via the following two channels: (1) the orbit
channel in which the charges with opposite momenta are decoupled via
Lorentz force in a magnetic field (i.e., the so-called orbital
limit), and (2) the spin channel is broken via the Zeeman effect in
which the spin singlet is transferred to spin triplet. At
sufficiently high magnetic field, superconductivity can be destroyed
by orbital and spin pair breaking. However, for a triplet
superconductor, the spin channel is robust against magnetic field
and, therefore, its upper critical field can be largely enhanced in
comparison with singlet superconductors \cite{Fujimoto}.

According to the Werthamer-Helfand-Hohenberg (WHH) method
\cite{WHH147}, the orbital limited upper critical field is given by:

\begin{equation}
\mu_0H_{c2}^{orb}(0)=-0.69T_c(d\mu_0H_{c2}/dT)_{T_c}. \label{eq:one}
\end{equation}

For a weak-coupling BCS superconductor, the Pauli paramagnetic
limiting field can be expressed as \cite{Clogston, Chandrasekhar}:

\begin{equation}
\mu_0H_{c2}^P(0)[\rm{Tesla}]=1.86T_c[\rm K]. \label{eq:two}
\end{equation}

For Y$_2$C$_3$, the superconducting transition temperature and the
initial slope of the upper critical field are $T_c^{TDO}$ = 15.5 K
and $(d\mu_0H_{c2}/dT)_{T_c}$ = -2.3 T/K, respectively. From Eq. (1)
and Eq. (2), one can then calculate the upper critical fields
associated with the orbital limit and the spin Pauli paramagnetism,
which gives $\mu_0H_{c2}^{orb}(0) = 24.5$T and $\mu_0H_{c2}^P(0) =
28.8$T. Therefore, our experimentally derived value of
$\mu_0H_{c2}^{TDO}(0) = 30$ T is larger than those obtained within
the BCS theory. Furthermore, $\mu_0H_{c2}(T_c)$ shows upturn
curvatures close to $T_c$ and also at low temperature, deviating
from the predictions of the WHH theory \cite{WHH147} and the
Ginzburg-Landau theory \cite{Tinkham} in which $\mu_0H_{c2}(T_c)$
usually gets flat as the temperature goes to zero. Both the large
upper critical field $\mu_0H_{c2}(0)$ and the upturn curvature in
$\mu_0H_{c2}(T_c)$ are typically found in non-centrosymmetric
superconductors with a large antisymmetric spin-orbital coupling
\cite{Kimura98, Okuda}. However, we still couldn't rule out the
possibilities that the upper critical field might be enhanced by
strong coupling. To find out whether these unique properties,
including the large upper critical field and its unusual temperature
dependence, are attributed to the spin-triplet state as a result of
ASOC effect in non-centrosymmetric superconductors, further
measurements are desired in order to elucidate the order parameter
symmetry. The recent NMR \cite{Harada} and $\mu$SR \cite{Kuroiwa}
measurements claimed that Y$_2$C$_3$ is a two-band superconductor
like in MgB$_2$ \cite{Nagamatsu}. However, these measurements have
been only performed down to 2K and its data resolution is poor at
low temperature. Thus it is difficult to fit their low temperature
behavior precisely. Further lower temperature measurements, e.g.,
the penetration depth, specific heat and NMR, are, therefore, highly
desired. One possible scenario may happen as follows: the
contributions of spin-triplet state and spin-singlet states are
comparable in Y$_2$C$_3$, nodes may only develop at very low
temperature and $H_{c2}$ can be enhanced, as we see here, due to the
contribution of spin-triplet state.

\section{Conclusion}

In summary, we have determined the upper critical field
$H_{c2}(T_c)$ of Y$_2$C$_3$ by measuring the magnetization and the
TDO-based resonant frequency. We found that Y$_2$C$_3$ posses a
large upper critical field of $\mu_0H_{c2}(0)$$\simeq$30 T,
exceeding the spin paramagnetic limit in weak coupling. Furthermore,
an upturn curvature is observed in $\mu_0H_{c2}(T_c)$ at low
temperatures. These unusual magnetic properties might highlight the
importance of broken inversion symmetry on the superconducting
properties of Y$_2$C$_3$, in which the contributions from
spin-triplet state might play a role.

\section{Acknowledgements}
We acknowledge the helpful discussion with M. B. Salamon. This work
was supported by NSFC(NO.10874146, NO.10934005), the National Basic
Research Program of China (NO.2009CB929104), the PCSIRT of the
Ministry of Education of China, Zhejiang Provincial Natural Science
Foundation of China and the Fundamental Research Funds for the
Central Universities. Work at NHMFL-LANL is performed under the
auspices of the National Science Foundation, Department of Energy
and State of Florida. J.A. was partially supported by "High-Tech
Research Center Project" for Private Universities and Grant-in-Aid
for Scientific Research from Ministry of Education, Culture, Sports,
Science and Technology, Japan.


\begin{thebibliography}{00}


\bibitem{Anderson59} P. W. Anderson, J. Phys. Chem. Solids \textbf{11}, (1959) 26.
\bibitem{Anderson84} P. W. Anderson, Phy. Rev. B \textbf{30}, (1984) 4000.
\bibitem{Gor'kov} L. P. Gor'kov and E. I. Rashba, Phys. Rev. Lett. \textbf{87}, (2001) 037004.
\bibitem{Yip} S. K. Yip, Phys. Rev. B \textbf{65}, (2002) 144508.
\bibitem{Frigeri04} P. A. Frigeri, D. F. Agterberg, A. Koga, and M. Sigrist, Phys. Rev. Lett. \textbf{92}, (2004) 097001.
\bibitem{Samokhin} K. V. Samokhin, E. S. Zijlstra, and S. K. Bose, Phys. Rev. B \textbf{69}, (2004) 094514.
\bibitem{Bauer04} E. Bauer, G. Hilscher, H. Michor, C. Paul, E. W. Scheidt, A. Gribanov, Y. Seropegin, H. Noel and M. Sigrist, and P. Rogl, Phys. Rev. Lett. \textbf{92}, (2004) 027003.
\bibitem{Yuan} H. Q. Yuan, D. F. Agterberg, N. Hayashi, P. Badica, D. Vandervelde, K. Togano, M. Sigrist and M. B. Salamon, Phys. Rev. Lett. \textbf{97}, (2006) 017006.
\bibitem{Yuan 08} H. Q. Yuan, M. B. Salamon, P. Badica, and K. Togano, Phys. B \textbf{403}, (2008) 1138.
\bibitem{Nishiyama} M. Nishiyama, Y. Inada, and G. Q. Zheng, Phys. Rev. Lett. \textbf{98}, (2007) 047002.
\bibitem{Takeya} H. Takeya, M. Elmassalami, S. Kasahara, and K. Hirata, Phys. Rev. B, \textbf{76}, (2007) 104506.
\bibitem{Bauer07} E. Bauer, H. Kaldarar, A. Prokofiev, E. Royanian, A. Amato, J. Sereni, W. Bramer-Escamilla, and I. Bonalde, J. Phys. Soc. Jpn. \textbf{76}, (2007) 051009.
\bibitem{Takeuchi} T. Takeuchi, T. Yasuda, M. Tsujimo, H. Shishido, R. Settai, H. Harima, and Y. Onuki, J. Phys. Soc. Jpn. \textbf{76}, (2007) 014702.
\bibitem{Bonalde} I. Bonalde,W. Br\"{a}mer-Escamilla, and E. Bauer, Phys. Rev. Lett.\textbf{94}, (2005) 207002.
\bibitem{Kimura76} N. Kimura, Y. Muro, and H. Aoki, J. Phys. Soc. Jpn. \textbf{76}, (2007) 051010.
\bibitem{Kimura98} N. Kimura, K. Ito, K. Saitoh, Y. Umeda, and H. Aoki, Phys. Rev. Lett. \textbf{98}, (2007) 197001.
\bibitem{Sugitani} I. Sugitani, Y. Okuda, H. Shishido, T. Yamada, A. Thamizhavel, E. Yamamoto, T. D. Matsuda, Y. Haga, T. Takeuchi, R. Settai, and Y. Onuki, J. Phys. Soc. Jpn. \textbf{75}, (2006) 043703.
\bibitem{Okuda} Y. Okuda, Y. Miyauchi, Y ida, Y. Takeda, C. Tonohiro, Chie, Y. Oduchi,T. Yamada, N. D. Dung, T. D. Matsuda, Y. Haga, T. Takeuchi, M. Hagiwara, K. Kindo, H. Harima, K. Sugiyama, R. Settai,and Y.¨­nuki, J. Phys. Soc. Jpn. \textbf{76}, (2007) 044708.
\bibitem{Mukuda} H. Mukuda, T. Fujii, T. Ohara, A. Harada, M. Yashima, Y. Kitaoka, Y. Okuda, R. Settai, and Y. Onuki, Phys. Rev. Lett.\textbf{100}, (2008) 107003.
\bibitem{Amano} G. Amano, S. Akutagawa, T. Muranaka, Y. Zenitani, and J. Akimitsu, J. Phys. Soc. Jpn. \textbf{73}, (2004) 530.
\bibitem{Akutagawa76} S. Akutagawa, and J. Akimitsu, J. Phys. Soc. Jpn. \textbf{76}, (2007) 024713.
\bibitem{Harada} A. Harada, S. Akutagawa, Y. Miyamichi, H. Mukuda1, Y. Kitaoka, and J. Akimitsu, J. Phys. Soc. Jpn. \textbf{76}, (2007) 023704.
\bibitem{Kuroiwa} S. Kuroiwa, Y. Saura, J. Akimitsu, M. Hiraishi, M. Miyazaki, K. H. Satoh, S. Takeshita, and R. Kadono, Phys. Rev. Lett. \textbf{100}, (2008) 097002.
\bibitem{Nakane04} T. Nakane, and T. Mochiku, Appl. Phys. Lett. \textbf{84}, (2004) 2859.
\bibitem{Fujimoto} S. Fujimoto, J. Phys. Soc. Jpn. \textbf{76}, (2006) 051008.
\bibitem{WHH147} N. R. Werthamer, E. Helfand, and P.C. Hohenberg, Phys. Rev.\textbf{147}, (1966) 295.
\bibitem{Clogston} A. M. Clogston, Phys. Rev. Lett. \textbf{9}, (1962)266.
\bibitem{Chandrasekhar} B. S. Chandrasekhar, J. Appl. Phys. Lett. \textbf{1}, (1962)7.
\bibitem{Tinkham} M. Tinkham, \emph{Introduction to Superconductivity (2nd ed.)}, McGrawHill, New York (1996).
\bibitem{Nagamatsu} J. Nagamatsu, N. Nakagawa, T. Muranaka, Y. Zenitani, and J. Akimitsu, Nature \textbf{410}, (2001) 63.

\end{thebibliography}
\end{document}